\documentclass[
 aps,
 pra,
 amsmath,amssymb,
 reprint,
superscriptaddress,
twocolumn,
]{revtex4-2}

\usepackage{graphicx}
\usepackage{dcolumn}
\usepackage{bm}
\usepackage{mathtools}
\usepackage{dsfont}
\usepackage{array}
        \newcolumntype{M}[1]{>{\centering\arraybackslash}m{#1}}

\usepackage[colorinlistoftodos]{todonotes}

\begin{document}


\title{Quantum Wavefront Correction via Machine Learning for Satellite-to-Earth CV-QKD}

\author{Nathan~K.~Long}
\email{nathan.long1@unsw.edu.au}
\affiliation{School of Electrical Engineering and Telecommunications, University of New South Wales, Kensington, NSW, Australia}
 \affiliation{Quantum Technology Group, Sensors and Effectors Division, Defence Science and Technology Group, Edinburgh, SA, Australia} 
\author{Ziqing~Wang}
\affiliation{School of Electrical Engineering and Telecommunications, University of New South Wales, Kensington, NSW, Australia}
\author{Benjamin~P.~Dix-Matthews}
\affiliation{International Centre for Radio Astronomy Research, University of Western Australia, Perth, WA, Australia} %
\author{Alex~Frost}
\affiliation{International Centre for Radio Astronomy Research, University of Western Australia, Perth, WA, Australia}
\author{John~Wallis}
\affiliation{International Centre for Radio Astronomy Research, University of Western Australia, Perth, WA, Australia}
\author{Kenneth~J.~Grant}
\affiliation{School of Electrical Engineering and Telecommunications, University of New South Wales, Kensington, NSW, Australia}
\author{Robert~Malaney}
\affiliation{School of Electrical Engineering and Telecommunications, University of New South Wales, Kensington, NSW, Australia}



\date{\today}

\begin{abstract}
State-of-the-art free-space continuous-variable quantum key distribution (CV-QKD) protocols use phase reference pulses to modulate the wavefront of a real local oscillator at the receiver, thereby compensating for wavefront distortions caused by atmospheric turbulence. It is normally assumed that the wavefront distortion in these phase reference pulses is identical to the wavefront distortion in the quantum signals, which are multiplexed during transmission. However, in many real-world deployments, there can exist a relative wavefront error (WFE) between the reference pulses and quantum signals, which, among other deleterious effects, can severely limit secure key transfer in satellite-to-Earth CV-QKD. In this work, we introduce novel machine learning-based wavefront correction algorithms, which utilize multi-plane light conversion for decomposition of the reference pulses and quantum signals into the Hermite-Gaussian (HG) basis, then estimate the difference in HG mode phase measurements, effectively eliminating this problem. Through detailed simulations of the Earth-satellite channel, we demonstrate that our new algorithm can rapidly identify and compensate for any relative WFEs that may exist, whilst causing no harm when WFEs are similar across both the reference pulses and quantum signals. We quantify the gains available in our algorithm in terms of the CV-QKD secure key rate. We show channels where positive secure key rates are obtained using our algorithms, while information loss without wavefront correction would result in null key rates.


\end{abstract}

\maketitle

\section*{Introduction}

The realization of fault-tolerant quantum computing has the potential to break current encryption systems~\cite{Cheng2021}, requiring the advent of a new security paradigm to combat quantum algorithms. Quantum key distribution (QKD) uses the principles of quantum mechanics to share secure keys between parties, offering unconditional information-theoretic security (e.g.~\cite{Grosshans2002}). In particular, the use of a low Earth orbit satellite and ground station network for a Quantum Internet could facilitate global security using QKD.

Satellite-Earth continuous-variable QKD (CV-QKD) involves encoding secret key information on the quadratures of the electric field of weak quantum signals followed by  transmission through the turbulent atmospheric channel. In order to measure the quantum signal at the receiver, a bright transmitted local oscillator (TLO) is multiplexed (usually in polarization) with the transmission so that distortion of the phase wavefront can be quantified and used for coherent measurement of the signal (e.g.~\cite{Villasenor2020}). However, the TLO can be attacked by an eavesdropper, making the coherent measurement process insecure. Therefore,  state-of-the-art CV-QKD protocols instead transmit reference pulses with the quantum signals, which are measured at the receiver, and then encode the measured information onto a real local oscillator (RLO) generated at the receiver~\cite{Marie2017,Wang2018_llo,Williams2024}. 

As a first approximation, it is assumed that the reference pulses and quantum signals have the same phase after passing through an atmospheric channel; however, previous work reveals that there could exist a phase error between them after homodyne/heterodyne detection~\cite{Marie2017, Kish2021, Long2025_phase} -  an issue discussed in detail later. Although one-dimensional phase errors in the time domain have been analyzed previously in this context, no work has explored spatial wavefront errors (WFEs) between the wavefronts of the reference pulses and quantum signals. This topic is the focus of our work.

Shack-Hartmann wavefront sensors are commonly used for optical wavefront detection; however, they are sensitive to non-uniform illumination~\cite{Mahe2000, Akondi2019}, and scintillation is often high in satellite-Earth channels (e.g.~\cite{Bufton1977}). As such, we instead explore the use of a multi-plane light converter (MPLC) for wavefront detection. MPLCs are capable of decomposing an electric field on any arbitrary basis - here we focus on Hermite-Gaussian (HG) modes (e.g., as in~\cite{Billault2021}). MPLCs are emerging as an alternative to adaptive optics in free-space optical communications~\cite{Cho2022}. Our work investigates the relative WFEs between the HG modes (hereafter referred to as ``relative-mode WFEs'') by simulating the decomposition of the HG modes at the receiver and analyzing the reference pulse and quantum signal measurements.


Machine learning (ML), having been shown to be useful in estimating phase errors in the one-dimensional time domain in CV-QKD~\cite{Long2023_survey, Huang2024}, offers a potential solution to estimate relative-mode WFEs. A neural network (NN) can be used to map the relationship between reference pulse-mode WFEs and signal-mode WFEs. The trained model can then take the reference pulse-mode WFEs as input and output an estimate of the signal-mode WFEs, without the need for assumptions-based physical models or computationally intensive numerical search algorithms. Here, we design transformer NNs (also termed intelligent wavefront correction algorithms) to estimate the signal-mode WFEs which is then used to reconstruct the original signal wavefront. This corrected wavefront is mapped onto an RLO to compensate for the relative WFE between the reference pulses and the quantum signals. The RLO is then mixed with the quantum signal, using balanced homodyne detection, to recover the secure key information encoded on the quantum signal quadratures. 

The main aim of our work is to demonstrate how our new intelligent wavefront correction algorithms can deliver improved secure key rates when differential error between the reference pulses and quantum signals is present, with no negative impact on the key rates otherwise. We argue that their inclusion in next-generation CV-QKD systems for the satellite-Earth channels will assist in the ongoing development of the Quantum Internet.



The rest of this work is presented as follows: a system overview for the CV-QKD protocol and MPLC hardware is given in Section~\ref{sec:sys}, a description of satellite-Earth phase screen simulations is given in Section~\ref{sec:channel}, the WFE theory is outlined in Section~\ref{sec:wfes}, a discussion of the NNs is given in Section~\ref{sec:tnn}, the NN estimation performance is shown in Section~\ref{sec:perf}, wavefront correction results are summarized in Section~\ref{sec:correction}, an analysis of the effects on secure key rates is presented in Section~\ref{sec:skrs}, and concluding remarks are made in Section~\ref{sec:concl}.

\section{System Overview} \label{sec:sys}

In the following exposition, the quantum signals are assumed to be in the form of weak pulses and the reference pulses in the form of strong pulses. However, it will help clarify our exposition to henceforth refer to the pulsed quantum signals simply as ``signals''. 
Although our intelligent wavefront correction algorithms can be applied both in the downlink and the uplink of satellite-Earth channels, here we will focus entirely on the downlink channel, the satellite-to-Earth channel. The Earth-to-satellite uplink channel will have poorer transmissivity, and quantum communication through it could be better served via teleportation~\cite{Villasenor2021}. 

\subsection{CV-QKD Protocol}

Our RLO-based CV-QKD protocol involves encoding secure key information on the signal quadratures on the satellite, which we call Alice (subscript $A$). The signals are polarization multiplexed with reference pulses, then transmitted across satellite-to-Earth channels, where it has previously been assumed that the reference pulse phase WFE $\Delta\Phi_R(x,y)$ experiences the same distortion across the atmospheric channel as the signal phase WFE $\Delta\Phi_S(x,y)$, ${\Delta\Phi_R(x,y) \approx \Delta\Phi_S(x,y)}$ for each pulse. Bob (subscript $B$) demultiplexes the reference pulses and signals at the receiver, then measures $\Delta\Phi_R(x,y)$ using a wavefront sensor and maps $\Delta\Phi_R(x,y)$ onto an RLO, which is used to measure the signal through homodyne detection. 

However, as discussed in more detail later (see Section~\ref{sec:wfes}) in real-world transmissions of multiplexed reference pulses and signals, the difference between $\Delta\Phi_R(x,y)$ and $\Delta\Phi_S(x,y)$ could be non-zero. Therefore, mapping an estimation of the true value of $\Delta\Phi_S(x,y)$ on the RLO could result in greater coherence between the RLO and the signal, than with the \textit{approximation} ${\Delta\Phi_R(x,y) \approx \Delta\Phi_S(x,y)}$, when measuring the quadrature values of the signal, leading to higher secure key rates through the channel.

\begin{figure}
    \begin{centering}
    \includegraphics[scale=0.87]{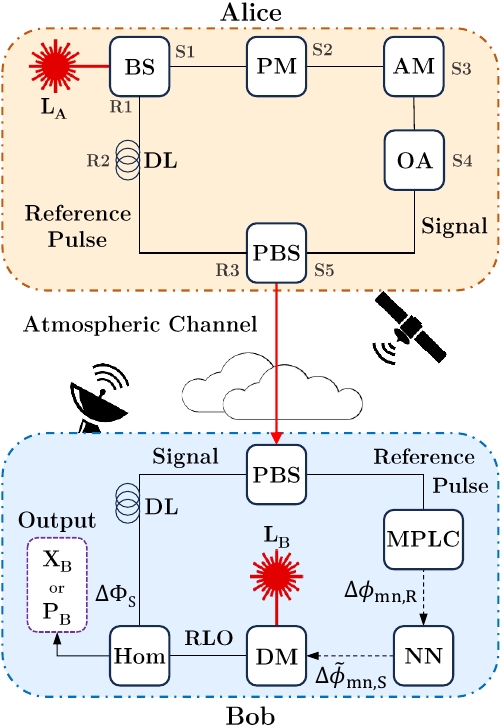}
    \caption{Schematic outlining the preparation and measurement of coherent states with our intelligent wavefront correction algorithm. L is a laser source, BS is a beam splitter, AM is an amplitude modulator, PM is a phase modulator, OA is an optical attenuator, PBS is a polarized beam splitter, MPLC is a multi-plane light converter, Hom is a homodyne detector, DM is a deformable mirror, DL is a delay line, and~NN is neural network. {Zernike WFEs are applied to the signals at S1-S5 and reference pulses at R1-R3. The quadrature measurements $X_B$ or $P_B$ contain the secure key information.}}\label{fig:circ}
    \end{centering}
\end{figure}

Although several wavefront sensing technologies exist, we focus on the use of an MPLC, which is capable of decomposing an electric field into weighted HG modes (or any other basis transformation). The reference pulse HG mode phase measurements in each mode $\Delta\phi_{mn,R}$, termed the reference pulse-mode WFEs, can then be used to \textit{estimate} the signal HG mode phase measurements in each mode $\Delta\phi_{mn,S}$, termed the signal-mode WFEs, where we construct NN-based intelligent wavefront correction algorithms to estimate $\Delta\phi_{mn,S}$ {(rather than just using $\Delta\phi_{mn,R}$ as an approximation of $\Delta\phi_{mn,S}$)}. Estimates $\Delta\Tilde{\phi}_{mn,S}$, termed the estimated signal-mode WFEs of $\Delta\phi_{mn,S}$, are then used to reconstruct an estimation of $\Delta\Phi_S(x,y)$, which is mapped on the RLO using a deformable mirror, as shown in our schematic of the CV-QKD protocol in Figure~\ref{fig:circ}.

In brief, the practical steps required to implement our intelligent wavefront correction algorithm, using the RLO-based CV-QKD protocol, are as follows. 

\textit{Training:} (1)~Transmit multiplexed reference pulses and signals across channel. (2)~Split reference pulses and signals at receiver.
    (3)~Measure reference pulses and signals independently using an MPLC.
    (4)~Train neural network with $\Delta\phi_{mn,R}$ and $\Delta\phi_{mn,S}$ measurements.
    
\textit{Operation:} (5)~Transmit multiplexed reference pulses and signals across the channel.
    (6)~Split reference pulses and signals at receiver.
    (7)~Measure \textit{only} reference pulses using an MPLC.
    (8)~Input $\Delta\phi_{mn,R}$ to NN, outputting estimates $\Delta\Tilde{\phi}_{mn,S}$.
    (9)~Use $\Delta\Tilde{\phi}_{mn,S}$ to map the estimated signal phase WFE on the RLO.
    (10)~Measure signal via balanced homodyne detection using the RLO.




For clarity, we outline the full nomenclature of WFE types in this work in Table~\ref{tab:phases}. Note that the mode WFEs $\Delta\phi_{mn,R}$ and $\Delta\phi_{mn,S}$ represent the phase measurements in each HG mode for the reference pulses and signals at the output of the MPLC, regardless of the type of WFE applied to the hardware before measurement (described in Section~\ref{sec:wfes}), while the estimated signal-mode WFEs $\Delta\Tilde{\phi}_{mn,S}$ represent the estimates produced by the intelligent wavefront correction algorithms.

\begin{table*} 
    \caption{Wavefront error nomenclature.}
    \label{tab:phases}
    \begin{ruledtabular}
    \begin{tabular}{M{3cm}lp{9cm}}
    \textrm{Symbol}&
    \textrm{Name}&
    \textrm{Definition}\\
    \colrule
        \vspace{0.015 in} $\Delta\Phi_R(x,y)$ & Reference pulse phase WFE & Difference between the reference pulse phase at $\{x,y\}$ at the laser source and receiver. \\ \hline
        \vspace{0.015 in} $\Delta\Phi_S(x,y)$ & Signal phase WFE & Difference between the signal phase at $\{x,y\}$ at the laser source and receiver. \\ \hline
        \vspace{0.015 in} $\Delta\phi_{mn}$ & HG mode phase & Generic HG mode difference between the measured phase and zero phase, as defined in Eq.~\ref{eq:c_mn}. Note that we replace $\Delta\phi_{mn}$ with $\Delta\phi_{mn,R}$, $\Delta\phi_{mn,S}$, $\Delta\theta_{mn,R}$, and $\Delta\theta_{mn,S}$ in Eq.~\ref{eq:c_mn} when calculating HG mode decompositions. \\ \hline
        \vspace{0.015 in} $\Delta\phi_{mn,R}$ & Reference pulse-mode WFEs & Difference between the reference pulse HG mode phase at the laser source and reference pulse MPLC output HG mode phase at the receiver (see Eq.~\ref{eq:c_mn}), independent of the type of WFEs applied at transmitter or receiver. \\ \hline
        \vspace{0.015 in} $\Delta\phi_{mn,S}$ & Signal-mode WFEs & Difference between the signal HG mode phase at the laser source and signal MPLC output HG mode phase at the receiver (see Eq.~\ref{eq:c_mn}), independent of the type of WFEs applied at transmitter or receiver. \\ \hline
        \vspace{0.015 in} $\Delta\Tilde{\phi}_{mn,S}$ & Estimated signal-mode WFEs & Corrected (see Section~\ref{sec:tnn}), independent of the type of WFEs applied at transmitter or receiver. \\ \hline
        \vspace{0.015 in} $\Delta\phi_{mn,R} - \Delta\phi_{mn,S}$ & Relative-mode WFEs & Difference between the reference pulse and signal MPLC output HG mode phase at the receiver. \\ \hline
        \vspace{0.015 in} $\Delta\Phi(\rho,\beta)$ & Zernike WFEs & 2nd- and 3rd-order WFEs caused by imperfect optical hardware at the transmitter, defined using Zernike polynomials (see Section~\ref{sec:transmit_wfes}). \\ \hline
        \vspace{0.015 in} $\Delta\theta_{mn,R}$ & Reference pulse transmit WFEs & Difference between the received reference pulse HG mode phase after Zernike WFEs are applied at the transmitter (without cross-leakage WFEs) (see Section~\ref{sec:transmit_wfes}). \\ \hline
        \vspace{0.015 in} $\Delta\theta_{mn,S}$ & Signal transmit WFEs & Difference between the received signal HG mode phase after Zernike WFEs are applied at the transmitter (without cross-leakage WFEs) (see Section~\ref{sec:transmit_wfes}). \\ \hline
        \vspace{0.015 in} $\Delta\theta_{mn,X}$ & Cross-leakage WFEs & Difference between the signal HG mode phase and the reference pulse HG mode phase caused by photon leakage from the reference pulses to signals, and including crosstalk from the MPLC (without reference pulse and signal transmit WFEs) (see Section~\ref{sec:xtalk_wfes}). \\ \hline
        \vspace{0.015 in} ${\mathrm{Var}(\Delta\phi_{mn,R} - \Delta\phi_{mn,S})}$ & Default variance & Variance of the relative-mode WFEs (see Section~\ref{sec:perf}). \\ \hline
        \vspace{0.015 in} ${\mathrm{Var}(\Delta\Tilde{\phi}_{mn,S} - \Delta\phi_{mn,S})}$ & Correction variance & Variance of the difference between the estimated signal-mode WFEs and signal-mode WFEs (see Section~\ref{sec:perf}). \\
    \end{tabular}
    \end{ruledtabular}
\end{table*}


\subsection{Multi-Plane Light Conversion}

An electric field can be described using any arbitrary spatial mode basis, where a beam passing through an atmospheric channel can be defined as a linear combination of the mode basis with a complex phase factor varying over time due to turbulence fluctuations in the channel. 

An MPLC is a device capable of demultiplexing an electric field into $N$ HG spatial modes. A series of phase masks, separated by free space, are used to transform the input electric field into $N$ output spots of the fundamental mode HG$_{00}$, which are then coupled to separate single-mode fibers (SMFs)~\cite{Cho2022}.

HG modes represent solutions to the paraxial wave equation~\cite{Siegman1986}, and are a function of the wavelength $\lambda$, the beam waist $w_0$, and the location along the optical axis of the waist, $z_0$. HG modes can be described using beam radius $w(z)$, calculated as,
\begin{equation}\label{eq:beam_rad}
    w(z) = w_0 \sqrt{1 + \frac{(z-z_0)^2}{z^2_R}},
\end{equation}

\noindent where the radius of the curvature of the beam wavefront is ${R(z) = (z-z_0) + z^2_R/(z-z_0)}$, and the Gouy phase is ${\psi(z) = \arctan([z-z_0]/z_R)}$, representing an additional phase per mode due to its deviation from a plane wave. Both are calculated using the Rayleigh range ${z_R = \pi w^2_0 / \lambda}$. The normalized non-astigmatic HG modes are defined as~\cite{Cho2022},
\begin{equation}\label{eq:hg_modes}
    \begin{split}
    & \mathrm{HG}_{mn}(x,y,z) = \frac{e^{-\frac{\rho^2}{w(z)^2}-i\left(\frac{k\rho^2}{2R(z)} + k(z-z_0) - (n+m+1)\psi(z)\right)}}{\sqrt{2^{n+m-1} \pi n! m!} w(z)} \\
    & \cdot H_m\left(\frac{\sqrt{2}x}{w(z)}\right) H_n\left(\frac{\sqrt{2}y}{w(z)}\right),
    \end{split}
\end{equation}

\noindent where the wavenumber ${k=2\pi/\lambda}$ and the radial distance $\rho$ is given by the Cartesian coordinates $\{x,y\}$ as ${\rho^2 = x^2 + y^2}$. The modes are spatially modulated using the Hermite polynomials $H_m$ (or $H_n$),
\begin{equation}\label{eq:hg_poly}
    H_m(x) = m! \sum^{\lfloor \frac{m}{2} \rfloor}_{j=0} \frac{(-1)^j}{j!(m-2j)} \frac{x^{m-2j}}{2^j},
\end{equation}

\noindent where $\lfloor \ \rfloor$ represents the floor function. The order of HG modes $H_{mn}$ is $m+n$, resulting in indices ${0,1,\cdots,N}$. 

Any optical beam $E(x,y)$ can then be expressed in terms of an infinite linear combination of orthogonal HG modes~\cite{Cho2022},
\begin{equation}\label{eq:e_hg}
    E(x,y) = \sum^\infty_{mn} a_{mn} \mathrm{HG}_{mn} e^{i\Delta\phi_{mn}} = \sum^\infty_{mn} c_{mn} \mathrm{HG}_{mn}.
\end{equation}

\noindent The HG mode complex coefficients $c_{mn}$ are computed as follows,
\begin{equation}\label{eq:c_mn}
    c_{mn} = a_{mn} e^{i \Delta\phi_{mn}} = \frac{\iint E(x,y) \mathrm{HG}^*_{mn}(x,y) dxdy}{\iint \mathrm{HG}_{mn}(x,y) \mathrm{HG}^*_{mn}(x,y) dxdy},
\end{equation}

\noindent with amplitude $a_{mn}$ and phase $\Delta\phi_{mn}$, where we generically define the phase here - later replacing $\Delta\phi_{mn}$ with $\Delta\phi_{mn,S}$, $\Delta\phi_{mn,R}$, $\Delta\theta_{mn,S}$, or $\Delta\theta_{mn,R}$ when simulating decomposition of the electric fields of the reference pulses and signals into the HG basis. Once each HG mode is coupled into a separate SMF, with the electric field $E_{mn}$ and scaled as $c_{mn}$, then the voltage of each mode $V_{mn}$ is measured using a photodetector, which is proportional to the power in each mode $P_{mn}$,
\begin{equation}\label{eq:mplc_meas}
    V_{mn} \propto P_{mn} = |E_{mn}|^2 = |c_{mn}|^2.
\end{equation}

\noindent Note that interferometry (e.g. heterodyne detection) of each mode would be required to measure $\Delta\phi_{mn}$.

\begin{figure}
    \begin{centering}
    \includegraphics[scale=0.57]{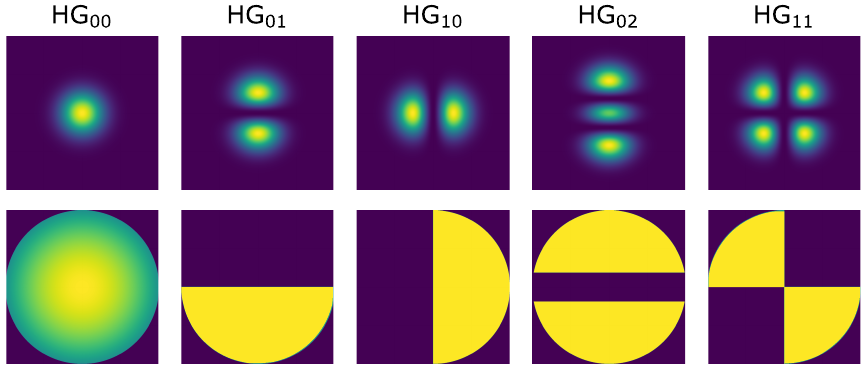}
    \caption{Hermite-Gaussian modes, with the transverse intensity profiles on the top row and the transverse phase profiles on the bottom row {(for an undefined channel)}.}\label{fig:hg_modes}
    \end{centering}
\end{figure}

Figure~\ref{fig:hg_modes} shows the first five HG modes, where the transverse intensity profiles of each mode are shown on the top row, while the corresponding transverse phase profiles are shown on the bottom row. In this work, we examine CV-QKD performance for three different numbers of modes: $N=10$, $N=30$, and $N=50$.

\section{Satellite-to-Earth Channel} \label{sec:channel}

We implemented a phase screen model to simulate the propagation of the reference pulses and signals through atmospheric channels from the satellite to the Earth, based on~\cite{Schmidt2010, Beck2016, Martin1988}. A phase screen model utilizing Monte Carlo simulations with a Fresnel split-step propagator is detailed in the work~\cite{Schmidt2010}. Essentially, a series of phase screens are used to represent a turbulent atmospheric volume by assuming that an optical beam travels as if in a vacuum until it interacts with a phase screen at its center. The phase screen then perturbs the phase of the beam wavefront as it passes through it. The methodology of~\cite{Schmidt2010} is implemented as a Python package in~\cite{Beck2016}.

The works~\cite{Schmidt2010} and~\cite{Beck2016} both assume that the atmosphere has homogeneous turbulence properties along the propagation path. Although this assumption may be reasonable for beam propagation across short horizontal distances, the atmosphere changes substantially when considering the vertical distances from a satellite to the Earth~\cite{Martin1988}.

For satellite-to-Earth channels, the distribution of phase screens is modified as the density of the atmosphere reduces with increased altitude~\cite{noaa1976}, resulting in different turbulence properties, where we base the satellite-to-Earth phase screen modeling on~\cite{Martin1988} (as implemented in~\cite{Wang2020}). The atmosphere is divided into $N_s$ slabs, bounded by the altitudes $h_j$ and $h_{j-1}$~in~meters, where $j$ increases in altitude from~1~to~$N_s$. The thickness of each slab $\Delta L_{j}$ is estimated as ${\Delta L_j = (h_j - h_{j-1})/\cos(\theta_z)}$, where $\theta_z$ is the zenith angle of the satellite. The ground station is at an altitude $h_0$ and the satellite is at an altitude $H$. 

The turbulence within each slab is defined by the scintillation index $\sigma^2_{I_j}$ and the Fried parameter $r_{0_j}$, where $\sigma^2_{I_j}$ for each slab is calculated as,
\begin{equation}\label{eq:sigmai}
    \sigma^2_{I_j} = \exp \left[ \frac{0.49 \sigma^2_{R_j}}{(1 + 1.11 \sigma^{12/5}_{R_j})^{7/6}} + \frac{0.51 \sigma^2_{R_j}}{(1 + 0.69 \sigma^{12/5}_{R_j})^{5/6}} \right] - 1,
\end{equation}

\noindent for the Rytov variance $\sigma^2_{R_j}$, calculated as a function of the altitude, $\theta_z$, $C^2_n$ and $k$,
\begin{equation}\label{eq:sigmar}
    \sigma^2_R = 2.25 k^{7/6} \sec^{11/6}(\theta_z) \int_{h_{j-1}}^{h_j} C^2_n (h)(h-h_0)^{5/6} dh.
\end{equation}

\noindent The altitude-dependent refractive index structure parameter $C^2_n(h)$ can be modeled as a function of the root-mean-square (RMS) wind speed $v_{rms}$~in~m/s and ground-level $C^2_n$ in m$^{-2/3}$, $C^2_n(0)$ as~\cite{Andrews2005},
\begin{equation}\label{eq:cn2}
    \begin{split}
        C^2_n(h) &= 0.00594 (v_{rms}/27)^2 \ (h \times 10^{-5})^{10} \ e^{(-h/1000)} \\ 
        & + 2.7 \times 10^{-16} \ e^{(-h/1500)} + C^2_n(0) e^{(-h/100)}.
    \end{split}
\end{equation}

\noindent The RMS wind speed is calculated as a function of the altitude-dependent wind profile $v(h)$,
\begin{equation}\label{eq:wind_rms}
    v_{rms} = \left[ \frac{1}{15 \times 10^3} \int_{5 \times 10^3}^{20 \times 10^3} v^2(h) \ dh \right]^{1/2},
\end{equation}

\noindent with $v(h)$ given in~m/s. We adopt the Bufton wind profile~\cite{Andrews2005} when simulating $v(h)$, calculated as,
\begin{equation}\label{eq:wind}
    v(h) = v_y + 30 \exp \left[ -\left( \frac{h - 9400}{4800} \right)^2 \right],
\end{equation}

\noindent where $v_y$ is the ground-level wind speed in~m/s. Note that the slew rate is not considered. 

\noindent The Fried parameter for each slab is then given by,
\begin{equation}\label{eq:r0_sat}
     r_{0_j} = \left[ 0.423k^2 \sec(\theta_z) \int_{h_{j-1}}^{h_j} C^2_n(h) dh \right]^{-3/5}.
\end{equation}

\noindent The distribution of phase screens for the satellite-to-Earth channel is optimized by determining $N_s$ and $h_j$ using a numerical search to solve for ${\sigma^2_{I_j} < 0.1}$ and ${\sigma^2_{I_j} < 0.1 \sigma^2_I}$. 

\begin{figure}
    \centering
    \includegraphics[scale=0.5]{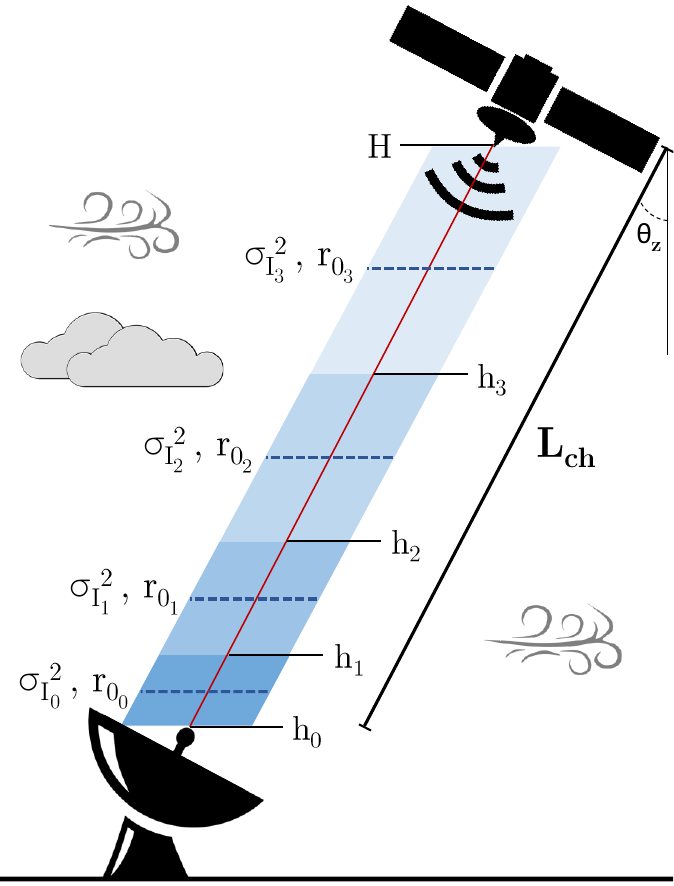}
    \caption{Satellite-to-Earth phase screen model for channel length $L_{ch}$ and azimuth angle $\theta_z$. Phase screens (dashed blue lines) simulate beam distortion using the scintillation index $\sigma^2_{I_j}$ and Fried parameter $r_{0_j}$, which are placed at the center of an atmospheric volume bounded by $h_{j-1}$ and $h_j$.}
    \label{fig:ps_model}
\end{figure}

Figure~\ref{fig:ps_model} gives an example of how phase screens are distributed, with the distance between phase screens decreasing as the atmosphere becomes denser, where $L_{ch}$ represents the total propagation distance from the transmitter to the receiver.

The parameters used in our simulation of the satellite-to-Earth link are given in Table~\ref{tab:sim_params}, where the inner scale~$l_0$ and outer scale~$L_0$ of turbulence are used to define turbulent eddy sizes in the phase screen model~\cite{Schmidt2010}. Note that we generate an independent $C^2_n(0)$ value (see range in Table~\ref{tab:sim_params}) for each instance of the phase screen simulations, thus the phase screen vertical distribution optimization is undertaken for each simulation. Further, we assume that birefringence is negligible across the channel.

\begin{table}[b]
    \caption{Simulation parameters.}
    \label{tab:sim_params}
    \begin{ruledtabular}
    \begin{tabular}{lc}
    \textrm{Parameter}&
    \textrm{Value}\\
    \colrule
        Satellite altitude ($H$) & 500 km \\ 
        Receiver altitude ($h_0$) & 2.00 km \\ 
        Outer scale of turbulence ($L_0$) & 5.00 m \\ 
        Inner scale of turbulence ($l_0$) & 0.025 m \\ 
        Satellite zenith angle ($\theta_z$) & $0^\circ$ \\ 
        Laser wavelength ($\lambda$) & 1550 nm \\ 
        Beam waist radius ($w_0$) & 0.150 m  \\ 
        Receiver radius ($R_r$) & 1.25 m \\ 
        RMS wind speed ($\nu_{rms}$) & 21.0 m/s \\ 
        Ground-level $C^2_n$ $\left(C^2_n(0)\right)$ & {$1.70 \times 10^{-15}$} - $10^{-14}$ m$^{-2/3}$  \\
    \end{tabular}
    \end{ruledtabular}
\end{table}

\section{Wavefront Errors} \label{sec:wfes}
In this Section, we discuss the different forms of WFE that can occur in our system due to imperfections in optical hardware and system calibration.

\subsection{Differential Wavefront Errors} \label{sec:transmit_diffwfes}

Although previous work has investigated phase errors between reference pulses and post-heterodyne/homodyne detection signals in RLO-based CV-QKD~\cite{Marie2017, Kish2021, Long2025_phase}, most have assumed that reference pulse phase WFEs equate to signal phase WFEs, ${\Delta\Phi_R(x,y) \approx \Delta\Phi_S(x,y)}$. Although this may be true in CV-QKD setups with near-perfect hardware and calibration, there are cases where ${\Delta\Phi_R(x,y) \neq \Delta\Phi_S(x,y)}$. Imperfections in optical hardware manufacturing, as well as post-manufacturing stresses and thermal fluctuations, can distort a wavefront passing through them~\cite{Doyle2000}. Imperfect reference pulse and signal polarization multiplexing can cause photon leakage from reference pulses to signals~\cite{Wang2018_llo, Huang2016}, and imperfect MPLC phase masks can lead to internal crosstalk between HG modes~\cite{Billault2021}. Although every CV-QKD setup is different, these imperfections can lead to ${\Delta\Phi_R(x,y) \neq \Delta\Phi_S(x,y)}$, which would affect CV-QKD performance (see Sections~\ref{sec:perf}~to~\ref{sec:skrs}). {Note that experimental support for differential WFEs has recently been shown in~\cite{Long2025_qce}.}

Additional differences in reference pulse and signal phase errors may occur through the channel. These issues are discussed in detail in~\cite{Marie2017}. In theory no phase error difference is caused by the turbulent channel, provided the complex wavelength structure (width, amplitude, and phase distributions of all wavelengths) for the strong reference pulses and weak signal pulses are identical, no leakage across the polarized multiplexed signals is present, and all optical hardware and calibration is perfect. In reality, it is unlikely that all these conditions are met exactly. 

\subsection{Transmitter Wavefront Errors} \label{sec:transmit_wfes}

We assume that the reference pulses and signals are Gaussian beams generated by the same laser source at the transmitter. The signals then pass through a beamsplitter (BS), phase modulator (PS), amplitude modulator (AM), optical attenuator (OA), and polarized beamsplitter (PBS) at the transmitter; while the reference pulses pass through a BS, delay line (DL), and PBS (see Figure~\ref{fig:circ}). Imperfections in optical hardware can result in distortion of a beam\textquoteright s wavefront~\cite{Doyle2000}. Given the different pieces of optical hardware through which the reference pulses and signals pass independently, their wavefronts could be distorted independently, resulting in a relative WFE between them at transmission.

For example, the properties of optical equipment can be affected by temperature gradients across the surfaces, absorption, coating materials, residual polishing compounds, and inhomogeneities in the optical material itself~\cite{Doyle2000}. As such, changes in the refractive index within the material, due to thermal inhomogeneities and the state of stress of optical materials, can cause distortions to a beam passing through them, resulting in WFEs. This effect could be particularly prominent on satellites given the large temperature fluctuations in space, as well as the stresses caused when launching a satellite into space.

Zernike polynomials are commonly used to describe the types of WFE induced by optical hardware, where we base the following theory on~\cite{Wyant1992}. Zernike polynomials are an infinite set of polynomials, defined by the polar coordinates $\rho$ and $\beta$ for a unit circle aperture as,
\begin{equation} 
    Z^p_q(\rho, \beta)=
    \begin{cases*}
        R^p_q(\rho) \cos(p\beta),        & \ \text{if $p\geq0$}, \\
        R^{-p}_q(\rho) \sin(-p\beta),    & \ \text{otherwise},
    \end{cases*}
\end{equation}

\noindent given the integers $p$ and $q$, where ${q - p}$ must be even, and ${q \geq p \geq 0}$. The radial polynomial term $R^p_q(\rho)$ is calculated as,
\begin{equation}\label{eq:zernike_poly}
  R^p_q(\rho) = \sum^{\frac{q-p}{2}}_{j=0} \frac{(-1)^j (q-j)!}{j! (\frac{p+q}{2} -j)! (\frac{p-q}{2} -j)!} \rho^{q-2j}.
\end{equation}

\begin{figure}
    \begin{centering}
    \includegraphics[scale=0.55]{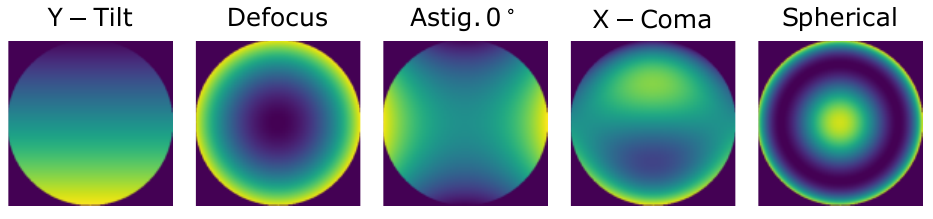}
    \caption{Wavefront errors due to imperfections in optical hardware: y-tilt, de-focus, 0$^\circ$ astigmatism, x-coma, and spherical.}\label{fig:zern}
    \end{centering}
\end{figure}

\noindent We focus on the second-order WFE \textit{tilt}; as well as the third-order WFEs \textit{defocus}, \textit{astigmatism}, \textit{coma}, and \textit{spherical} (shown in Figure~\ref{fig:zern}), defined by their Zernike polynomials in Table~\ref{tab:zernike}. The Zernike WFEs $\Delta\Phi(\rho, \beta)$ can then be calculated in terms of the Zernike polynomials using the coefficient $a^p_q$,
\begin{equation}\label{eq:zernike_phase}
  \Delta\Phi(\rho, \beta) = a^p_q Z^p_q(\rho, \beta),
\end{equation}

\noindent where $0 \leq \rho \leq 1$ and $0 \leq \beta \leq 2\pi$.

\begin{table}[b]
    \caption{Zernike coefficients for wavefront errors.}
    \label{tab:zernike}
    \begin{ruledtabular}
    \begin{tabular}{cccl}
    $n$ & $m$ & \textrm{Zernike Polynomial} & \textrm{Error Type} \\
    \colrule
        1 & -1 & $\rho\sin(\beta)$ & Y-tilt ($Z^{-1}_1$) \\ 
        2 & 0 & $2\rho^2-1$ & De-focus ($Z^{0}_2$) \\ 
        2 & 2 & $\rho^2 \cos 2 \beta$ & Astigmatism 0$^\circ$ ($Z^{2}_2$) \\ 
        2 & -2 & $\rho^2 \sin 2 \beta$ & Astigmatism 45$^\circ$ ($Z^{-2}_2$) \\ 
        3 & 1 & $(3\rho^3-2\rho) \cos \beta$ & X-coma ($Z^{1}_3$) \\ 
        3 & -1 & $(3\rho^3-2\rho) \sin \beta$ & Y-coma ($Z^{-1}_3$) \\ 
        4 & 0 & $6\rho^4-6\rho^2 + 1$ & Spherical ($Z^{0}_4$) \\ 
    \end{tabular}
    \end{ruledtabular}
\end{table}

WFE specifications for optical hardware are often given in terms of their RMS error, calculated as ${\mathrm{RMS} = (\frac{1}{J} \sum_j \Delta\phi^2_j)^{1/2}}$ for $J$ discrete error measurements $\Delta\phi_j$ across the discretized wavefront. Therefore, for known Zernike WFE types (such as given in Table~\ref{tab:zernike}), which are defined by their RMS error by the manufacturers, we can solve Eq.~\ref{eq:zernike_phase} to calculate $a^p_q$ so that $\Delta\Phi(\rho, \beta)$ can be derived for each Zernike WFE at the transmitter.

The Zernike WFEs ${\Delta\Phi_{S1}(\rho,\beta) \ \mathrm{to} \ \Delta\Phi_{S5}(\rho,\beta)}$ are applied to the signals and the Zernike WFEs ${\Delta\Phi_{R1}(\rho,\beta) \ \mathrm{to} \ \Delta\Phi_{R4}(\rho,\beta)}$ are applied to the reference pulses as they pass through each piece of hardware at the transmitter, as shown in Figure~\ref{fig:circ}. The reference pulses and signals are then sent across the channel, as simulated by the phase screen model described in Section~\ref{sec:channel}. 

The RMS errors used to generate each of the Zernike WFEs are given in Table~\ref{tab:sources_dphi}, where they approximate ``precise'' equipment, as defined by~\cite{meetoptics2025}. Note that $\Delta\Phi_{R2}(\rho,\beta)$ represents a phase shift throughout the wavefront caused by an imperfectly calibrated DL, {added to account for the difference in paths taken by the reference pulses and signals}.

\begin{table}[b]
    \caption{Transmitter wavefront error types and parameterization.}
    \label{tab:sources_dphi}
    \begin{ruledtabular}
    \begin{tabular}{clc}
    \textrm{Source} & \textrm{WFE Type} & \textrm{WFE Value} \\
    \colrule
        $\Delta\Phi_{S1}(\rho,\beta)$ & Astigmatism 0$^\circ$ & 0.081$\lambda$ (RMS) \\ 
        $\Delta\Phi_{S2}(\rho,\beta)$ & Spherical & 0.100$\lambda$ (RMS) \\ 
        $\Delta\Phi_{S3}(\rho,\beta)$ & Defocus & 0.078$\lambda$ (RMS) \\ 
        $\Delta\Phi_{S4}(\rho,\beta)$ & X-coma & 0.091$\lambda$ (RMS) \\
        $\Delta\Phi_{S5}(\rho,\beta)$ & Y-tilt & 0.096$\lambda$ (RMS) \\ 
        $\Delta\Phi_{R1}(\rho,\beta)$ & Y-coma & 0.106$\lambda$ (RMS) \\ 
        $\Delta\Phi_{R2}(\rho,\beta)$ & Phase Shift & 0.869~rad \\ 
        $\Delta\Phi_{R3}(\rho,\beta)$ & Astigmatism 45$^\circ$ & 0.098$\lambda$ (RMS) \\ 
    \end{tabular}
    \end{ruledtabular}
\end{table}

After passing the reference pulse and signal beams through the satellite-to-Earth channel, their distorted electric fields are deconstructed into $N$ HG modes using Eq.~\ref{eq:c_mn}. The resulting phase components of the complex coefficients $c_{mn}$ for both the reference pulses and signals are defined by $\Delta\theta_{mn,R}$ (termed the reference pulse transmit WFEs) and $\Delta\theta_{mn,S}$ (termed the signal transmit WFEs) after propagating through the atmosphere.


\subsection{Crosstalk and Leakage Wavefront Errors} \label{sec:xtalk_wfes}

Apart from second- and third-order WFEs at the transmitter, additional hardware imperfections can cause relative-mode WFEs between the reference pulses and signals. Imperfect polarization multiplexing can lead to photon leakage from brighter reference pulses into weaker signals~\cite{Wang2018_llo, Huang2016}. In addition, crosstalk in the HG mode can occur due to imperfections in the design of the MPLC phase masks (e.g.~\cite{Billault2021}).

Imperfect polarization multiplexing would occur at the transmitter, where the transmit WFEs could leak from the reference pulses to the signals, which would then propagate over the atmosphere, before entering the MPLC at the receiver. The crosstalk between the MPLC HG modes is usually quantified in terms of the ratio of powers at the input to output of the MPLC~\cite{Billault2021}, extracted from the real component of the coefficients $c_{mn}$, rather than the phase component (see Eq.~\ref{eq:c_mn}). Imperfections in polarization multiplexing and phase-mask design within an MPLC will vary for each CV-QKD set-up, leading to unique additional WFEs. However, as an example, we introduce a sinusoidal harmonic function to represent the additional WFEs in each mode, termed the cross-leakage WFEs $\Delta\theta_{mn,X}$, given by,
\begin{equation}\label{eq:ddphi_mn_2}
    \Delta\theta_{mn,X} = \sum^3_{d=1} A_{{mn,d}} \sin(S_{{MN}} \cdot \Theta_{{MN,R}} ) ,
\end{equation}


\noindent {where the subscript $X$ represents the \textit{cross}-leakage}, $d$~represents the harmonic number, $A_{{mn,d}}$ is a vector of amplitudes for each harmonic for each mode (defined in Appendix~\ref{ap:a}), $S_{{MN}}$ is a vector of coefficients for each mode, and ${\Theta_{{MN,R}} = [\Delta\theta_{00,R}, \Delta\theta_{01,R}, \cdots, \Delta\theta_{mn,R}]^T}$. We calculate the vectors $S_{{MN}}$ by approximating the MPLC power ratio crosstalk matrices in~\cite{Billault2021}, as described in Appendix~\ref{ap:a}. We further note the analogy between HG modes and harmonic oscillators~\cite{Nienhuis1993}, as well as the discussion of phase correction for harmonic distortions (e.g.~\cite{Silva2018}), in our choice of the cross-leakage WFE function, while emphasizing that the additional WFE function could be different for every CV-QKD setup.

\section{Transformer Neural Network} \label{sec:tnn}

We build our intelligent wavefront correction algorithm architectures using transformer NNs (TNNs). TNNs are capable of processing an entire input vector (here, a vector of $\Delta\phi_{mn,R}$ for a single pulse) simultaneously, where patterns in the spatial relationship between all elements in the vector are derived and weighted, before being mapped to the desired output vector (here, a vector of $\Delta\phi_{mn,S}$ for a single pulse). We give a brief description of TNNs in this Section; refer to~\cite{Vaswani2017} for a more in-depth discussion of TNNs.

\begin{figure}
    \begin{centering}
    \includegraphics[scale=0.90]{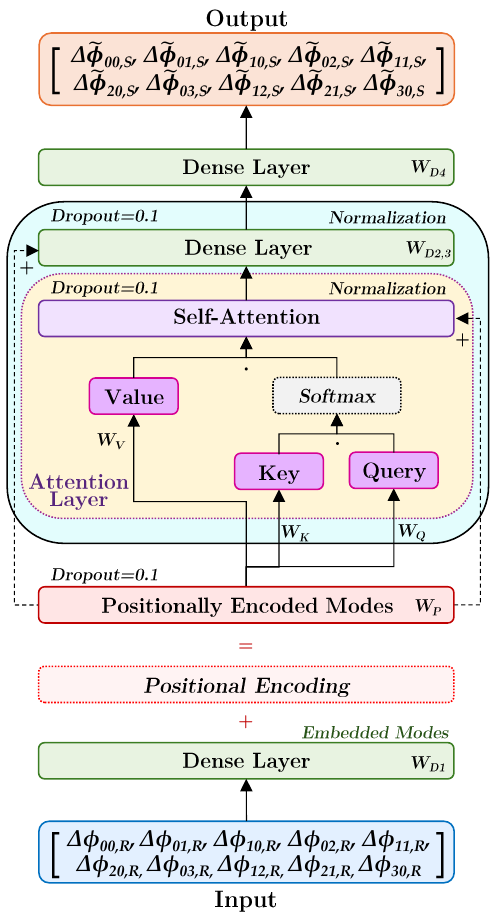} 
    \caption{TNN architecture, with an input and output of $N=10$ modes. The transformer layer architecture is shown as the light blue block, and the attention head architecture is shown in the yellow block. The weights $W_{D1}$, $W_{D2}$, $W_{D3}$, $W_{D4}$ represent the weights for the dense layers, while the weights $W_K$, $W_Q$, and $W_V$ represent the weights for the key, query, and value matrices, respectively.}\label{fig:tnn}
    \end{centering}
\end{figure}

The input vector is first embedded in the TNN using a (fully connected) \textit{dense layer}, resulting in a vector of length $d_m$. The positional encoding is then performed on each embedded mode, where their relative spatial positions are weighted, then summed with the embedded modes (shown at the bottom of Figure~\ref{fig:tnn}). The \textit{positionally encoded modes} (PEMs), a matrix of dimension $N \times d_m$, are then passed into the transformer layers (represented as the light blue block in Figure~\ref{fig:tnn}), which are made up of multiple \textit{attention heads} (shown as the yellow block in Figure~\ref{fig:tnn}) and two dense layers.

An attention head calculates three different matrices of weighted values, used to represent similarities between PEM elements. A \textit{query matrix} and \textit{key matrix} are calculated for each PEM, {providing ``context'' and ``relevance'', respectively,} then combined by taking the dot product between them. The dot product is then passed to a Softmax function~\cite{Wang2018_softmax}. The output of the Softmax function is combined with a separate \textit{value matrix} (calculated for the PEMs) via dot product, then summed with the original PEMs to obtain the ``similarity" between them, outputting a \textit{self-attention vector}. The key, query and value matrices are of dimension ${n_{attn} \times |d_m/n_{attn}|}$, where $n_{attn}$ is the number of attention heads in the TNN architecture. The self-attention vector has dimensions $d_m$.

Using multiple attention heads allows different relationships between the PEMs to be derived, improving the accuracy of the output estimations. The self-attention vector is then passed into a dense layer of dimension $d_t$ and another dense layer of dimension $d_m$, where they are again summed with the PEMs (helping to preserve positionality). The results from the transformer layers are then passed into a final dense layer, resulting in a vector of dimension $N$, before outputting a vector of estimates $\Delta\Tilde{\phi}_{mn,S}$ (shown for $N=10$ in Figure~\ref{fig:tnn}).




\subsection{Architectures}

We develop three different TNN architectures for the three different numbers of HG modes measured ($N=10$, $N=30$, and $N=50$), with their hyperparameters given in Table~\ref{tab:tnn_layers}.

\begin{table}[b]
    \caption{TNN architecture hyperparameters for $N=10$, $N=30$, and $N=50$ HG modes.}
    \label{tab:tnn_layers}
    \begin{ruledtabular}
    \begin{tabular}{lccc}
    \textrm{Hyperparameter} & $N=10$ & $N=30$ & $N=50$  \\
    \colrule
        Matrix dimension $d_{m}$ & 64 & 64 & 96 \\  
        Matrix dimension $d_{t}$ & 256 & 256 & 256 \\ 
        Attention heads $n_{attn}$ & 6 & 6 & 10 \\ 
        Transformer layers $\#$ & 3 & 4 & 4 \\
        Trainable parameters $\#$ & 147,037 & 195,985 & 339,545 \\ 
        Dropout & 0.10 & 0.10 & 0.10 \\
    \end{tabular}
    \end{ruledtabular}
\end{table}

The number of transformer layers, the number of attention heads, and the dimensions of the matrix ($d_{m}$ and $d_{t}$) vary between the architectures. An additional transformer layer is added for $N=30$ and $N=50$, due to the additional information being processed by the TNNs, while four additional attention heads are added for $N=50$. The dimension of the matrix $d_m$ was increased from 64~to~96 to account for the increase in the vector length of $N=50$ that is mapped through the TNN. In general, the number of training parameters in the TNN architectures increases as $N$ increases to map the increase in information from the input data $\Delta\phi_{mn,R}$ to the output data $\Delta\phi_{mn,R}$. Note that the hyper-parameterization of each TNN is achieved through an iterative process, aiming to achieve sufficient estimation performance, while limiting the number of trainable parameters.

\subsection{Training}

We simulate~24,000 instances of reference pulse and signal propagation through the satellite-to-Earth channels, followed by HG mode decomposition, which are split into 19,200~training instances and 4,800~testing instances. 

During training, the known input values $\Delta\phi_{mn,R}$ are assigned to the known $\Delta\phi_{mn,S}$ values. The weights $W_{D1}$, $W_{D2}$, $W_{D3}$, $W_{D4}$ represent the weights applied to the connections for dense layers, while the weights $W_K$, $W_Q$, and $W_V$ represent the weights applied to the key, query, and value matrices. All weights are optimized using the Adam optimizer~\cite{Bock2019} to minimize an RMS error loss function, training for~3,000 epochs using a batch size of~32. After training, all weight values remain fixed.

Further, we implement \textit{dropout} for the PEM, self-attention, and dense layers within the transformer layers. Dropout helps prevent overfitting the TNN weights by replacing a percentage of data within each matrix with zeros, improving the robustness of the intelligent wavefront correction algorithms~\cite{Srivastava2014}. The self-attention and dense layers, within the transformer layers, are also normalized to improve estimation performance~\cite{Ba2016}.

During operation, the TNNs take known input $\Delta\phi_{mn,R}$ measurements for each pulse, then output the estimates $\Delta\Tilde{\phi}_{mn,S}$ of $\Delta\phi_{mn,S}$ for that pulse. Estimates $\Delta\Tilde{\phi}_{mn,S}$ can then be used to reconstruct an estimated signal electric field ${\Tilde{E}_{mn,S}(x,y)}$ (see Section~\ref{sec:recon}), which can then be mapped onto the RLO to measure the signal with greater coherence.







\section{Estimation Performance} \label{sec:perf}

To analyze the performance of the intelligent wavefront correction algorithms in estimating $\Delta\phi_{mn,S}$, we calculate the \textit{correction variance} between $\Delta\Tilde{\phi}_{mn,S}$ and $\Delta\phi_{mn,S}$, where a perfect estimation performance would result in ${\mathrm{Var}(\Delta\Tilde{\phi}_{mn,S} - \Delta\phi_{mn,S}) = 0}$. We benchmark the estimation performance against the reference pulse approximation by calculating the \textit{default variance} between $\Delta\phi_{mn,R}$ and $\Delta\phi_{mn,S}$. When condition ${\mathrm{Var}(\Delta\Tilde{\phi}_{mn,S} - \Delta\phi_{mn,S}) < \mathrm{Var}(\Delta\phi_{mn,R} - \Delta\phi_{mn,S})}$ is achieved, the estimates $\Delta\Tilde{\phi}_{mn,S}$ give a better representation of $\Delta\phi_{mn,S}$ than $\Delta\phi_{mn,R}$, which leads to improved CV-QKD performance (shown in Section~\ref{sec:skrs}). 

 We present four different cases of relative-mode WFEs to demonstrate the effects on CV-QKD performance, using the transmit and cross-leakage WFEs described in Section~\ref{sec:wfes}. Case~0 represents when there are no transmit or cross-leakage WFEs, such that the approximation $\Delta\phi_{mn,R} \approx \Delta\phi_{mn,S}$ is true. Case~1 represents when there are only transmit WFEs, but no cross-leakage WFEs. Case~2 represents when there are transmit WFEs and cross-leakage WFEs. Case~3 represents when there are only cross-leakage WFEs, but no transmit WFEs.



\subsection{Case 0}

In the case of zero relative-mode WFEs, the approximation ${\Delta\phi_{mn,R} \approx \Delta\phi_{mn,S}}$ is true. We term this Case~0, representing CV-QKD systems with perfect optical hardware. To show that the intelligent wavefront correction algorithms will \textit{not} negatively affect CV-QKD performance for Case~0, TNNs are effectively trained to estimate ${\Delta\Tilde{\phi}_{mn,S} \approx \Delta\phi_{mn,R} \approx \Delta\phi_{mn,S}}$.

For Case~0, the default variances are zero, ${\mathrm{Var}(\Delta\phi_{mn,R} - \Delta\phi_{mn,S}) = 0}$, for all modes. Likewise, the correction variances approach zero, ${\mathrm{Var}(\Delta\Tilde{\phi}_{mn,S} - \Delta\phi_{mn,S}) \rightarrow 0}$, with the maximum and minimum variances for $N=10$, $N=30$, and $N=50$ given in Table~\ref{tab:var_c0}.

\begin{table}[b]
    \caption{Minimum and maximum correction variances ${\mathrm{Var}(\Delta\Tilde{\phi}_{mn,S} - \Delta\phi_{mn,S})}$ for~$N=10$,~$N=30$, and~$N=50$ in Case~0.}
    \label{tab:var_c0}
    \begin{ruledtabular}
    \begin{tabular}{ccc}
    $N$ & Min. Correction Variance & Max. Correction Variance \\
    \colrule
        10 & 4.29 $\times$ 10$^{-8}$ rad$^2$ & 1.24 $\times$ 10$^{-6}$ rad$^2$ \\ 
        30 & 3.63 $\times$ 10$^{-5}$ rad$^2$ & 2.21 $\times$ 10$^{-4}$ rad$^2$ \\ 
        50 & 6.34 $\times$ 10$^{-4}$ rad$^2$ & 1.67 $\times$ 10$^{-3}$ rad$^2$ \\ 
    \end{tabular}
    \end{ruledtabular}
\end{table}

As expected, the intelligent wavefront correction algorithms can learn when there are no relative-mode WFEs, outputting the estimate ${\Delta\Tilde{\phi}_{mn,S} \approx \Delta\phi_{mn,R}}$. Therefore, a CV-QKD setup with near-perfect hardware would not be disadvantaged by implementing the intelligent wavefront correction algorithms. However, if the hardware is disturbed or deteriorates over time, resulting in relative-mode WFEs, then the intelligent wavefront correction algorithms would be able to adapt and correct for them (as we show in the following analysis).

\subsection{Case 1}

In the case of the relative-mode WFEs only being associated with imperfections of the optics at the transmitter (i.e. the transmit WFEs), termed Case~1, the relative-mode WFEs output from the MPLC are given by,
\begin{equation}\label{eq:ddphi_c1}
    \Delta\phi_{mn,R} - \Delta\phi_{mn,S} = \Delta\theta_{mn,R} - \Delta\theta_{mn,S}.
\end{equation}

\begin{figure*}
    \begin{centering}
    \includegraphics[scale=0.57]{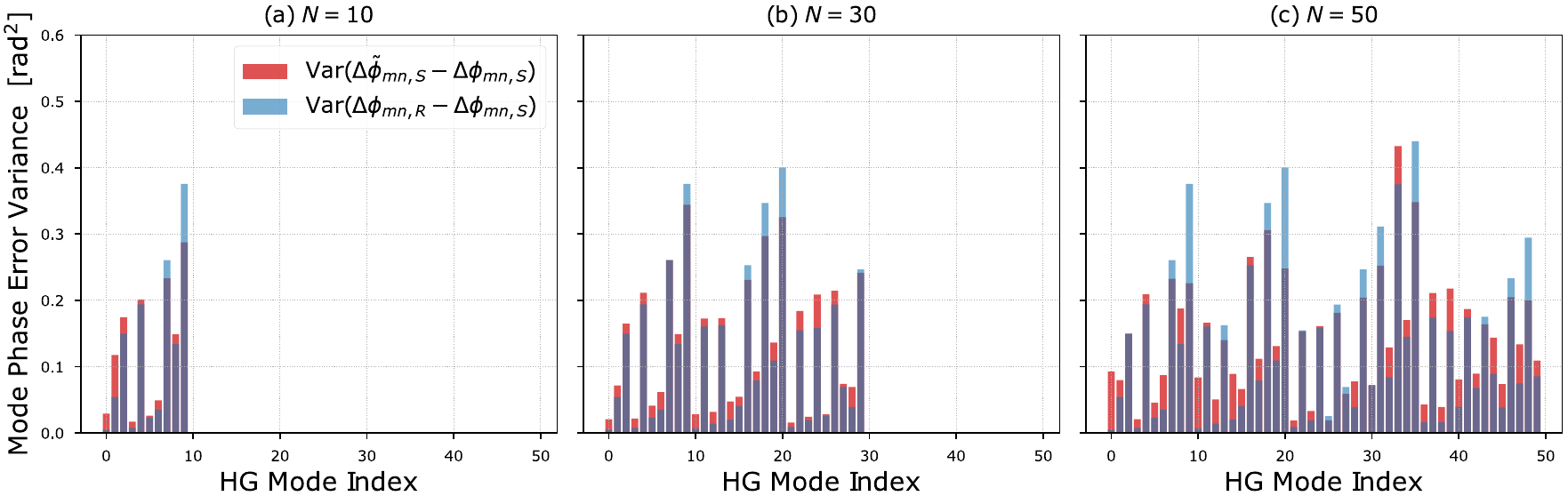} 
    \caption{Correction variance ${\mathrm{Var}(\Delta\Tilde{\phi}_{mn,S} - \Delta\phi_{mn,S})}$ (red bars) versus default variance ${\mathrm{Var}(\Delta\phi_{mn,R} - \Delta\phi_{mn,S})}$ (blue bars) for (a)~$N=10$, (b)~$N=30$, and (c)~$N=50$ in Case~1. {The purple sections show where the two variances overlap.}}\label{fig:dphi_var_c1}
    \end{centering}
\end{figure*}

Figure~\ref{fig:dphi_var_c1} shows the correction variance (red bars) and default variance (blue bars) for (a)~$N=10$, (b)~$N=30$, and (c)~$N=50$, while the purple sections show where the two variances overlap.

The default variances are lower than the correction variances for some modes (where the red bars are greater than the blue bars), while for other modes the correction variances are lower than the default variances (where the blue bars are greater than the red bars), seen for each value of $N$. In general, there is no large difference between the default variances and the correction variances. Although there are more modes where the default variances are lower than the correction variances, the modes where the correction variances are lower show a greater variance reduction than the default variances. In general, the greater reduction in the correction variances compensates for the greater number of modes in which the default variances are lower.

The estimation performance of the intelligent wavefront correction algorithms could be attributed to the phase wavefront perturbations caused by the channel (using the phase screen model~\cite{Schmidt2010}). The transmitted reference pulse and signal wavefronts (after Zernike WFEs are applied) are affected by Gaussian random processes, resulting in $\Delta\phi_{mn,R}$ - $\Delta\phi_{mn,S}$ for each mode being Gaussian distributions. However, while the correction variances do not provide a more accurate representation of $\Delta\phi_{mn,S}$ than the default variances, they do not provide a considerably worse representation. As such, incorporating our intelligent wavefront correction algorithms into CV-QKD protocols should not reduce CV-QKD performance when relative-mode WFEs are purely random.

\subsection{Case 2}

In the case of the relative-mode WFEs being associated both with imperfections of the optics at the transmitter (transmit WFEs) \textit{and} also with imperfections causing photon leakage and MPLC crosstalk (cross-leakage WFEs) - termed Case~2 - the relative-mode WFEs output from the MPLC are given by,
\begin{equation}\label{eq:ddphi_c2}
    \Delta\phi_{mn,R} - \Delta\phi_{mn,S} = \Delta\theta_{mn,R} - \Delta\theta_{mn,S} - \Delta\theta_{mn,X}.
\end{equation}

\begin{figure*}
    \begin{centering}
    \includegraphics[scale=0.59]{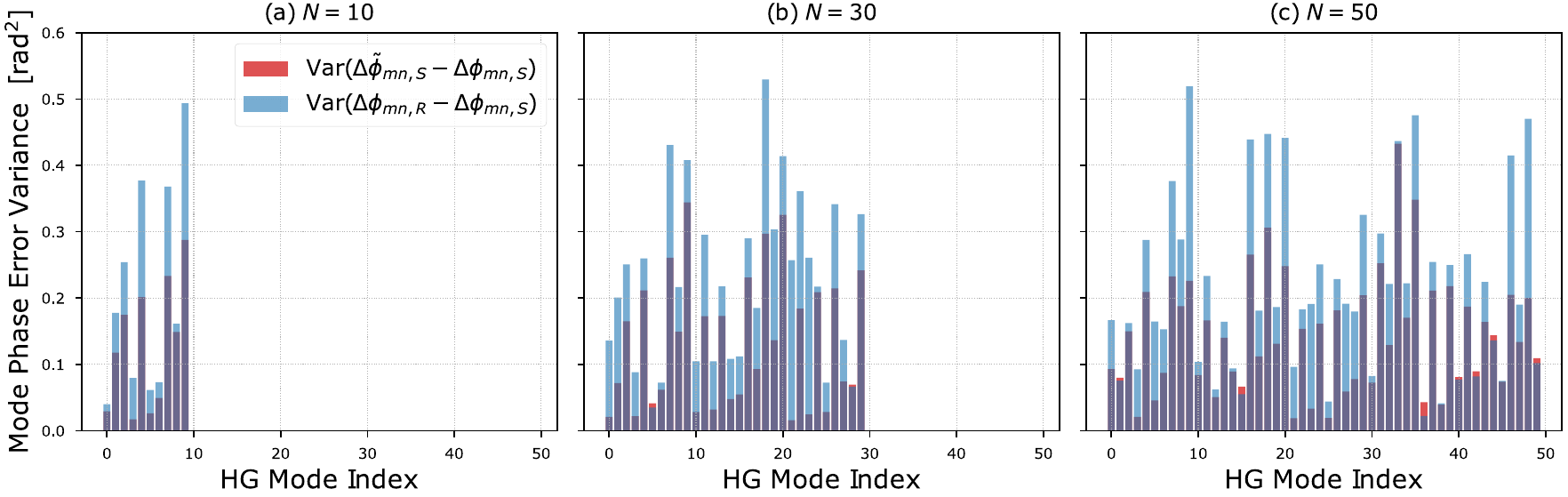} 
    \caption{Correction variance ${\mathrm{Var}(\Delta\Tilde{\phi}_{mn,S} - \Delta\phi_{mn,S})}$ (red bars) versus default variance ${\mathrm{Var}(\Delta\phi_{mn,R} - \Delta\phi_{mn,S})}$ (blue bars) for (a)~$N=10$, (b)~$N=30$, and (c)~$N=50$ in Case~2. {The purple sections show where the two variances overlap.}}\label{fig:dphi_var_c2}
    \end{centering}
\end{figure*}

As in Case~1, Figure~\ref{fig:dphi_var_c2} plots the correction variance (red bars) and the default variance (blue bars) for (a)~$N=10$, (b)~$N=30$, and (c)~$N=50$.

For almost all modes, the correction variances are lower than the default variances (where the blue bars are greater than the red bars). The differences between the correction variances and the default variances in Case~2 are also generally greater than the differences in Case~1, with default variances even exceeding 0.50~rad$^2$ (mode~18 for $N=30$ and mode~9 for $N=50$), after introducing cross-leakage WFEs.

The cross-leakage WFEs make $\Delta\phi_{mn,S}$ a function of $\Delta\phi_{mn,R}$, so that the intelligent wavefront correction algorithms can learn the relationship between them. Although the introduced cross-leakage WFEs will not represent every CV-QKD setup, the results for Case~2 highlight the utility of our intelligent wavefront correction algorithms - when a CV-QKD setup induces a relationship between $\Delta\phi_{mn,S}$ and $\Delta\phi_{mn,R}$, then estimation of $\Delta\phi_{mn,S}$ leads to improved CV-QKD performance (shown in Sections~\ref{sec:correction}~and~\ref{sec:skrs}). In addition, TNNs are particularly useful when there exists a relationship between the individual $\Delta\phi_{mn,S}$ and $\Delta\phi_{mn,R}$ modes. 




\subsection{Case 3}

In the case of the relative-mode WFEs being associated only with hardware imperfections causing photon leakage and MPLC crosstalk (cross-leakage WFEs) - termed Case~3 - the relative-mode WFEs output from the MPLC are given by,
\begin{equation}\label{eq:ddphi_c3}
    \Delta\phi_{mn,R} - \Delta\phi_{mn,S} = - \Delta\theta_{mn,X}.
\end{equation}

\begin{figure}
    \begin{centering}
    \includegraphics[scale=0.59]{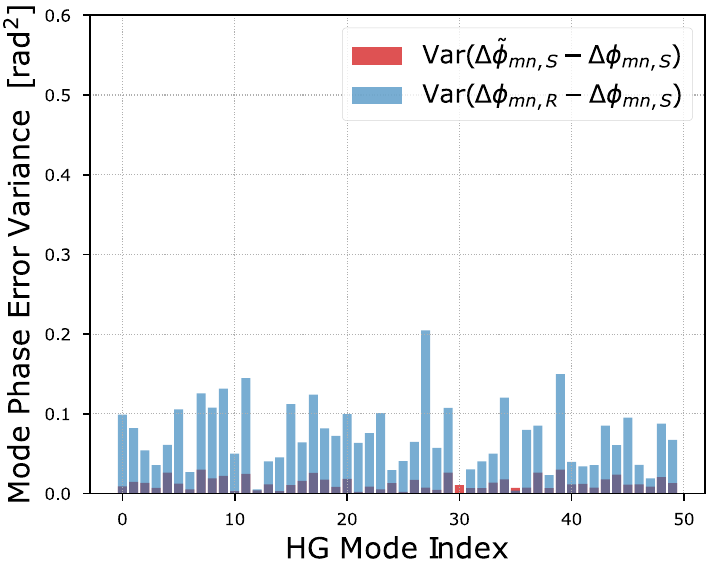} 
    \caption{Correction variance ${\mathrm{Var}(\Delta\Tilde{\phi}_{mn,S} - \Delta\phi_{mn,S})}$ (red bars) versus default variance ${\mathrm{Var}(\Delta\phi_{mn,R} - \Delta\phi_{mn,S})}$ (blue bars) for $N=50$ in Case~3. The purple sections show where the two variances overlap.}\label{fig:dphi_var_c3}
    \end{centering}
\end{figure}

\noindent {Overall, the default variances are lower for Case~3 than Case~1 or Case~2, with the correction variances being even lower, where the intelligent wavefront correction algorithms output estimates $\Delta\Tilde{\phi}_{mn,S}$ with lower correction variances than the default variances (as in Case~2). We provide an example of the estimation performance in Figure~\ref{fig:dphi_var_c3}, which plots the correction variance (red bars) and the default variance (blue bars) for $N=50$.} However, we assume that Case~2 is more representative than Case~3 in the presence of WFEs, given the noisier representation of the relative-mode WFEs in Eq.~\ref{eq:ddphi_c2} than for Eq.~\ref{eq:ddphi_mn_2}. For this reason, and given that the correction variances for Case~1 generally do not outperform the default variances, we focus on Case~2 when analyzing wavefront correction (Section~\ref{sec:correction}) and subsequent performance on secure key rates (Section~\ref{sec:skrs}) for CV-QKD.

\section{Wavefront Correction} \label{sec:correction}

\subsection{Wavefront Reconstruction} \label{sec:recon}

Any electric field can theoretically be reconstructed using an infinite number of HG modes (see Eq.~\ref{eq:e_hg}); however, given that MPLCs measure a limited number of modes ($N$), the wavefront reconstructions are imprecise.

We consider three scenarios to understand the performance of the $\Delta\Tilde{\phi}_{mn,S}$ estimations for Case~2. First, we reconstruct the electric field of the signal using the ``perfect'' MPLC measurement case, where the signal HG coefficients are used to reconstruct the signal electric field, 
$${E}_{mn,S}(x,y) = \sum^N_{mn} a_{mn,S} \mathrm{HG}_{mn} e^{i\Delta\phi_{mn,S}},$$

\noindent representing the greatest amount of information we could theoretically extract from the MPLC about the signal. Next, we consider the scenario where the reference pulse HG coefficients are used to reconstruct the electric field,
$${E}_{mn,R}(x,y) = \sum^N_{mn} a_{mn,R} \mathrm{HG}_{mn} e^{i\Delta\phi_{mn,R}},$$

\noindent which represents the approximation of ${E_{R}(x,y) \approx E_S(x,y)}$. Finally, we reconstruct the electric field using the estimates $\Delta\Tilde{\phi}_{mn,S}$, 
$$\Tilde{E}_{mn,S}(x,y) = \sum^N_{mn} a_{mn,R} \mathrm{HG}_{mn} e^{i\Delta\Tilde{\phi}_{mn,S}}.$$

\noindent Note that the reference pulse amplitudes $a_{mn,R}$ are used in the reconstruction using estimates $\Delta\Tilde{\phi}_{mn,S}$.

\begin{figure}
    \begin{centering}
    \includegraphics[scale=0.72]{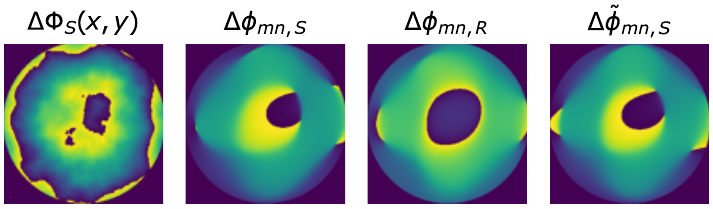} 
    \caption{True signal wavefront $\Delta\Phi_{S}(x,y)$ compared to wavefront reconstructions using $\Delta\phi_{mn,S}$, $\Delta\phi_{mn,R}$, and $\Delta\Tilde{\phi}_{mn,S}$ for $N=30$ modes.}\label{fig:mplc_recon}
    \end{centering}
\end{figure}

Figure~\ref{fig:mplc_recon} gives an example of the wavefront reconstruction for $\Delta\phi_{mn,S}$, $\Delta\phi_{mn,R}$, and $\Delta\Tilde{\phi}_{mn,S}$ using $N=30$ modes, as well as the true signal phase WFE $\Delta\Phi_S(x,y)$. It can be seen that the estimated signal wavefront reconstruction $\Tilde{E}_{mn,S}(x,y)$ more closely matches the signal wavefront reconstruction using $E_{mn,S}(x,y)$ than the reference pulse wavefront reconstruction using $E_{mn,R}(x,y)$.

\subsection{Coherent Efficiency} \label{sec:gamma}

To quantify the difference between the true signal electric field $E_S(x,y)$ and reconstructed electric fields $E_{mn}(x,y)$, we calculate the coherent efficiency $\gamma$ as~\cite{Liu2016},
\begin{equation}\label{eq:gamma_orig}
    \gamma = \frac{|\frac{1}{2} \iint_{\mathcal{D}_R}\left(E_S^* E_{mn} + E_SE^*_{mn}\right) \ dr|^2}{\left( \iint_{\mathcal{D}_R} |E_S|^2 \ dr\right) \left(\iint_{\mathcal{D}_R} |E_{mn}|^2 \ dr\right)},
\end{equation}

\noindent across the aperture with diameter $\mathcal{D}_R$ {for the radial integral $dr$}. A coherent efficiency of one represents perfect mode matching between the electric fields, e.g. ${E_S(x,y) = E_{mn}(x,y)}$, while a coherent efficiency of zero represents complete decoherence.

Table~\ref{tab:gamma} gives the results for the average coherent efficiencies for each of the three reconstruction scenarios presented, where we replace $E_{mn}(x,y)$ with $E_{mn,S}(x,y)$, $E_{mn,R}(x,y)$, or $\Tilde{E}_{mn,S}(x,y)$ in Eq.~\ref{eq:gamma_orig}.

\begin{table}[b]
    \caption{Coherent efficiencies between electric fields of true signal $E_{S}(x,y)$ and reconstructed electric fields of $E_{mn,S}(x,y)$, $E_{mn,R}(x,y)$, and $\tilde{E}_{mn,S}(x,y)$ for different numbers of HG modes $N$.}
    \label{tab:gamma}
    \begin{ruledtabular}
    \begin{tabular}{ccc}
    $N$ & $E_{mn}(x,y)$ & \textrm{Avg. Coherent Efficiency} \\
    \colrule
        10 & $E_{mn,S}(x,y)$ & 0.329 \\ 
        10 & $E_{mn,R}(x,y)$ & 0.229 \\ 
        10 & $\tilde{E}_{mn,S}(x,y)$ & 0.313 \\ \hline
        30 & $E_{mn,S}(x,y)$ & 0.507 \\ 
        30 & $E_{mn,R}(x,y)$ & 0.321 \\ 
        30 & $\tilde{E}_{mn,S}(x,y)$ & 0.484 \\ \hline
        50 & $E_{mn,S}(x,y)$ & 0.553 \\ 
        50 & $E_{mn,R}(x,y)$ & 0.349 \\ 
        50 & $\tilde{E}_{mn,S}(x,y)$ & 0.523 \\
    \end{tabular}
    \end{ruledtabular}
\end{table}

Firstly, comparing the perfect signal reconstructions $E_{mn,S}(x,y)$ with the reference pulse reconstructions $E_{mn,R}(x,y)$, we see that the coherent efficiency decreases from $\gamma=0.329$ to $\gamma=0.229$ for $N=10$, $\gamma=0.507$ to $\gamma=0.321$ for $N=30$, and from $\gamma=0.553$ to $\gamma=0.349$ for $N=50$. The estimated signal reconstructions $\tilde{E}_{mn,S}(x,y)$ show coherent efficiencies more closely matched to the signal wavefront reconstructions, with $\gamma=0.329$ to $\gamma=0.313$ for $N=10$, $\gamma=0.507$ to $\gamma=0.484$ for $N=30$, and from $\gamma=0.553$ to $\gamma=0.523$ for $N=50$, thus showcasing the benefit of our intelligent wavefront correction algorithm when there is a relationship between $\Delta\phi_{mn,S}$ and $\Delta\phi_{mn,R}$. The effects of coherent efficiency on secure key rates in CV-QKD follow in Section~\ref{sec:skrs}.

\section{Secure Key Rates} \label{sec:skrs}

In CV-QKD, signals are affected by excess noise contributions from channel noise $\xi_{ch}$ and detector noise $\xi_{det}$, both of which reduce the secure key rates attainable across a channel~\cite{Hosseinidehaj2019}. In this work, we adopt the GG02 CV-QKD protocol with reverse reconciliation~\cite{Grosshans2002}. Note that we apply the same security analysis as the entanglement-based GG02 protocol (given in~\cite{Laudenbach2018}). This means we assume that the ground station is trusted and an eavesdropper (Eve) does not have knowledge of the training process, nor is even aware that our new system is deployed at the ground station. That is, we take the reduced excess noise that our system can deliver to be a consequence of the correction to relative-mode WFEs between the reference pulses and signals. A formal security proof of the impact of our system when Eve is aware of its presence (and can interfere with the training process) is clearly a substantial task and should be the subject of future work. We also note that we have ignored finite key effects, so our secure key rates are valid only in the limit of large block lengths. Conversion of rates into bits/s would require a full reconciliation analysis, including the LDPC decoding timescale for decoding large block-length LDPC codes. This reconciliation will likely be the bottleneck for all CV-QKD protocols - the impact of our NN-runtime would be insignificant in comparison.

The coherent efficiency between the corrected RLO and signal contributes to the detector noise, where~\cite{Wang2019} derives the detector efficiency as a function of coherent efficiency as,
\begin{equation}\label{eq:xi_det}
    \xi_{det} = \frac{((1-\gamma) + \xi_{el}) \eta_{det}}{\gamma},
\end{equation}

\noindent for electronic noise $\xi_{el}$ and detector efficiency $\eta_{det}$. The detector noise is used to calculate the shared information between Alice and Bob $I_{AB}$, whereby increasing the coherent efficiency reduces $\xi_{det}$, increasing the shared information as follows~\cite{Wang2019}. The shared information is calculated as,
\begin{equation}\label{eq:inf_ab}
    I_{AB} = \frac{1}{2} \log_2(1 + \frac{T_f V_{mod}}{1 + T_f \xi_f}),
\end{equation}

\noindent for the modulation variance $V_{mod}$. Shared information is a function of transmissivity $T$, calculated as $T =(P_R/P_T) \eta_{det}$, where $P_T$ and $P_R$ represent the total power at the transmitter and receiver, respectively. Here, we apply an \textit{effective transmissivity} and an \textit{effective excess noise} term $T_f \xi_f$ to evaluate the secure key rates, given by,
\begin{equation}\label{eq:tf}
   T_f \xi_f = \xi_{ch} \langle T\rangle + \langle \xi_{det} \rangle + \left(\langle T \rangle - \langle \sqrt{T}\rangle^2\right) V_{mod}.
\end{equation}

\noindent The average transmissivity $\langle T \rangle = 0.652$ is calculated using the signal propagation through the phase screen simulations, while the average detector noise $\langle \xi_{det} \rangle$ is calculated using the coherent efficiencies described in Section~\ref{sec:gamma} for $N=10$, $N=30$, and $N=50$. The noise and efficiency parameters used to calculate the secure key rates are given in Table~\ref{tab:noise_params}.

\begin{table}[b]
    \caption{CV-QKD noise parameters.}
    \label{tab:noise_params}
    \begin{ruledtabular}
    \begin{tabular}{lc}
    \textrm{Parameter} & \textrm{Value} \\
    \colrule
        Detector efficiency ($\eta_{det}$) & 95$\%$~\cite{Kish2021}  \\ 
        Reverse reconciliation efficiency ($\beta_{r}$) & 95$\%$~\cite{Kish2021} \\ 
        Detector electronic noise ($\xi_{el}$) & 0.010 SNU~\cite{Wang2018} \\ 
        Channel noise ($\xi_{ch}$) & 0.0172~\cite{Kish2021} \\
    \end{tabular}
    \end{ruledtabular}
\end{table}

The covariance matrix $M_{AB}$ for the GG02 protocol, using homodyne detection of the signal, is given by~\cite{Wang2019, Laudenbach2018},
\begin{align}\label{eq:cov_mat}
        &M_{AB} = \begin{pmatrix}a\mathds{1} & c\sigma_z\\c\sigma_z & b\mathds{1}\end{pmatrix} \\
        &= \begin{pmatrix}(V_{mod}+1)\mathds{1} & \sqrt{T_f(V^2_{mod} + 2 V_{mod})} \sigma_z \\\sqrt{T_f(V^2_{mod} + 2 V_{mod})} \sigma_z & (T_f V_{mod} + 1 + \xi_{ch}) \mathds{1} \end{pmatrix}, \notag
\end{align}

\noindent where $\sigma_z$ is the Pauli-z matrix and $\mathds{1}$ is the identity matrix of dimension $2\times2$.

Holevo information $\chi_{BE}$ is the amount of information about a quantum state $\rho_{AB}$ attainable by Eve, calculated as $\chi_{BE} = S_{AB} - S_{A|B}$, for the von Neumann entropy $S_{AB}$ of the quantum state accessible by Eve and $S_{A|B}$ is the entropy after homodyne measurement by Bob~\cite{Laudenbach2018}. The subscripts $A$, $B$, and $E$ represent Alice, Bob, and Eve, respectively. If Eve has a purification of Alice and Bob\textquoteright s shared quantum state, $S(\rho_{AB}) = \sum_i g(\nu_i)$ for the function $g(x)$, given by,
\begin{equation}\label{eq:gx}
    g(x) = \frac{x+1}{2} \log_2 \left(\frac{x+1}{2}\right) - \frac{x-1}{2} \log_2 \left(\frac{x-1}{2}\right).
\end{equation}

\noindent The first two symplectic eigenvalues $\nu_{1,2}$ (the modulus of any eigenvalue) of $M_{AB}$ are derived as $\nu_{1,2} = \frac{1}{2}(z \pm [b - a])$, where $z = ([a+b]^2 - 4c^2)^{1/2}$~\cite{Weedbrook2012}. A third symplectic eigenvalue $\nu_{3}$ is calculated for the conditional matrix $M_{A|B}$, found by,
\begin{align}\label{eq:cov_mat_meas}
        \tilde{M}_{A|B} &= i\Omega M_{A|B} \\
        &= i\begin{pmatrix}0 & 1 \\ -1 & 0\end{pmatrix} \begin{pmatrix}a - \frac{c^2}{b} & 0 \\ 0 & a\end{pmatrix} = i\begin{pmatrix}0 & a \\ -a + \frac{c^2}{b} & 0\end{pmatrix}, \notag
\end{align}

\noindent which leads to $\nu_{3} = (a[a - \frac{c^2}{b}])^{1/2}$~\cite{Laudenbach2018}. The Holevo information can be derived as ${\chi_{BE} = g(\nu_1) + g(\nu_2) - g(\nu_3)}$.

Secure key rates $R_{sec}$ are calculated from shared information and Holevo information, $R_{sec} = \beta_r I_{AB} - \chi_{BE}$, with $\beta_r$ representing reverse reconciliation efficiency. Our secure key rate analysis assumes the asymptotic limit of the signal under a collective attack.

\begin{figure}
    \begin{centering}
    \includegraphics[scale=0.86]{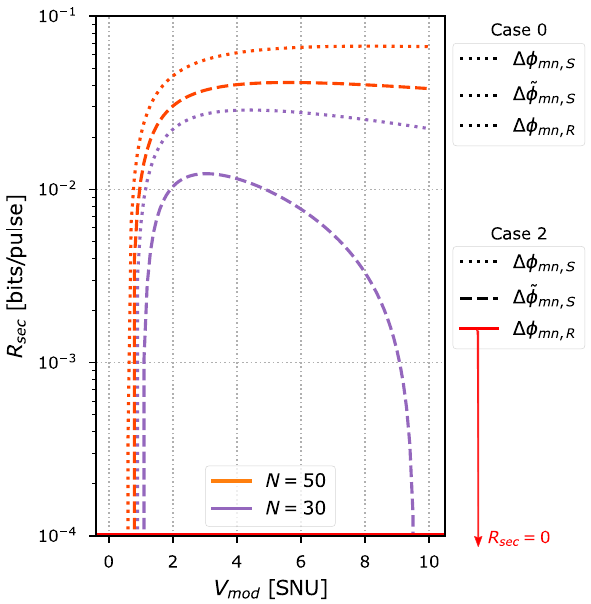} 
    \caption{Secure key rates versus modulation variance, {measured in shot noise units (SNU)}, for $N=30$ (purple) and $N=50$ (orange) modes. Both Case~0 and Case~2 results are presented, where ${\Delta\Tilde{\phi}_{mn,S} \approx \Delta\phi_{mn,R} \approx \Delta\phi_{mn,S}}$ results in equivalent secure key rates for Case~0, {such that the dotted lines represent the secure key rates attained for the $\Delta\Tilde{\phi}_{mn,S}$, $\Delta\phi_{mn,R}$, and $\Delta\phi_{mn,S}$ wavefront reconstructions. In Case~2, the secure key rates attained for the wavefront reconstructions using $\Delta\Tilde{\phi}_{mn,S}$ are plotted as dashed lines and $\Delta\phi_{mn,S}$ are plotted as dotted lines, while null key rates are attained for $\Delta\phi_{mn,R}$}. The curve for $\Delta\Tilde{\phi}_{mn,S}$ at $N=30$ is caused by the trade-off between shared information and excess noise as $V_{mod}$ increases.}\label{fig:skrs}
    \end{centering}
\end{figure}

Figure~\ref{fig:skrs} plots the secure key rates for Case~0 and Case~2 against modulation variance for the wavefront reconstructions outlined in Section~\ref{sec:recon}, where the secure key rates for $N=50$ (colored orange) are higher than for $N=30$ (colored purple). Note that null key rates are achieved for $N=10$.

\begin{figure*}[t]
    \begin{centering}
    \includegraphics[scale=0.60]{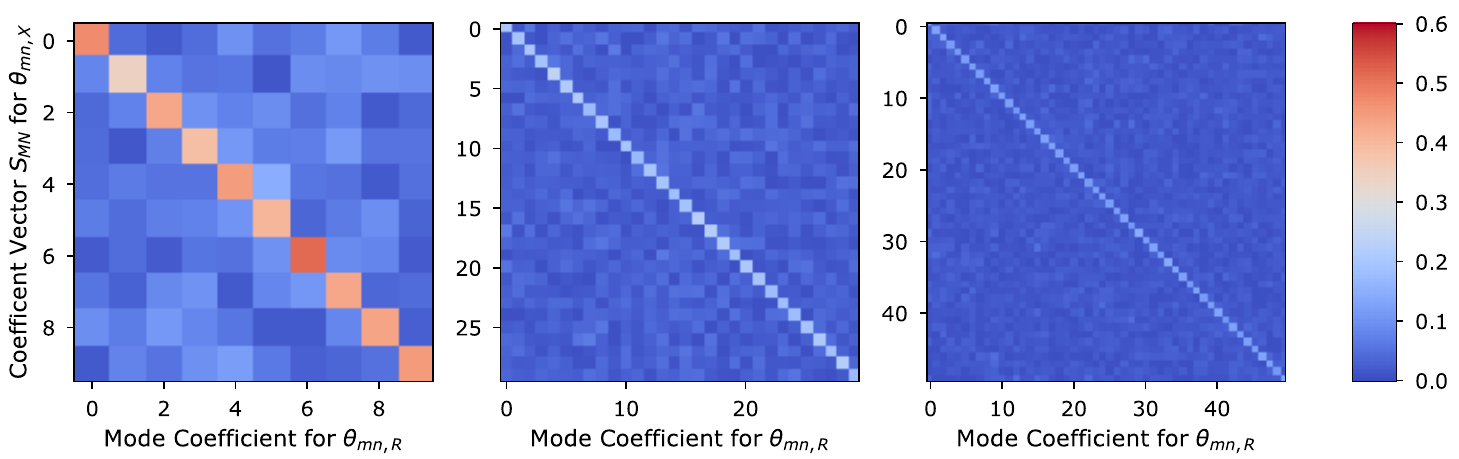}
    \caption{Cross-leakage WFE coefficient matrices of vectors $S_{{MN}}$ for (a)~$N=10$, (b)~$N=30$, and (c)~$N=50$.}\label{fig:mplc_xtalk}
    \end{centering}
\end{figure*}

For Case~0, where ${\Delta\Tilde{\phi}_{mn,S} \approx \Delta\phi_{mn,R} \approx \Delta\phi_{mn,S}}$, secure key rates will be equivalent for estimations $\Delta\Tilde{\phi}_{mn,S}$ and approximations $\Delta{\phi}_{mn,R}$ (represented as dotted lines for $N=30$ and $N=50$), where shared information is the highest attainable for $N$ modes, given equal coherent efficiencies for $\Delta\Tilde{\phi}_{mn,S}$, $\Delta{\phi}_{mn,R}$ and $\Delta{\phi}_{mn,S}$. As such, if the assumption $\Delta\phi_{mn,R} \approx \Delta\phi_{mn,S}$ is true (optical hardware is high precision and perfectly calibrated, {resulting in negligible relative-mode WFEs}), then the intelligent wavefront correction algorithms output $\Delta\Tilde{\phi}_{mn,S} \approx \Delta\phi_{mn,R}$, attaining equivalent secure key rates.

Importantly, when considering Case~2, null key rates are found for the $\Delta\phi_{mn,R}$ reconstructions when $\Delta\phi_{mn,R} \neq \Delta\phi_{mn,S}$ for the channel and hardware specifications outlined in this work. The decrease in shared information, resulting from the increased detector noise, is caused by the low coherent efficiencies between the signal wavefront and the reconstructed reference pulse wavefront. However, after our intelligent wavefront correction algorithms are implemented, positive key rates are achieved (dashed lines) for $N=30$ and $N=50$. Here, we exemplify the impact of our intelligent wavefront correction algorithms - when large WFEs occur in a CV-QKD setup, then the intelligent wavefront correction algorithms could prove crucial in enabling practical CV-QKD, providing a step forward in establishing a realizable Quantum Internet.

\section{Conclusion} \label{sec:concl}

We presented an analysis of wavefront errors associated with a real local oscillator-based continuous-variable quantum key distribution protocol across satellite-to-Earth channels, including the potential harm wavefront errors could have on secure key rates. Focusing on the use of a multi-plane light converter for wavefront sensing, we designed intelligent wavefront correction algorithms to correct for wavefront errors, increasing secure key rates across the simulated channels, thereby increasing the practical feasibility of implementing a satellite-based quantum communications network.

\begin{acknowledgments}
The Commonwealth of Australia (represented by the Department of Defence) supports this research through a Defence Science Partnerships agreement. 
\end{acknowledgments}

\appendix


\section{Cross-Leakage Wavefront Error Derivation} \label{ap:a}

The cross-leakage WFE for each mode is a function of the amplitudes $A_{{mn,d}}$, the coefficient vector $S_{{MN}}$, and the reference pulse transmit WFEs ${\Theta_{{MN,R}} = [\Delta\theta_{00,R}, \Delta\theta_{01,R}, \cdots, \Delta\theta_{mn,R}]^T}$. Therefore, the cross-leakage WFE for each mode is a function of \textit{all} reference pulse transmit WFE modes, weighted by a coefficient (see Eq.~\ref{eq:ddphi_mn_2}). We extract the coefficient vectors $S_{{MN}}$ from the matrices shown in Figure~\ref{fig:mplc_xtalk} for (a)~$N=10$, (b)~$N=30$, and (c)~$N=50$, {which approximate the MPLC crosstalk matrices in~\cite{Billault2021}.}

The vertical axis represents the cross-leakage WFE mode to which the coefficient vector $S_{{MN}}$ is applied, while the horizontal axis represents the specific coefficients applied to the modes in $\Theta_{{MN,R}}$. As can be seen, the diagonal coefficients are greater than the off-diagonal ones. This is implemented as we assume that more leakage would occur from $\Delta\theta_{mn,R}$ to $\Delta\theta_{mn,S}$ in the same mode, where $mn$ for $\Delta\theta_{mn,S}$ equals $mn$ for $\Delta\theta_{mn,R}$, while less leakage would occur when $mn \neq mn$, as is true when considering MPLC power ratio crosstalk~\cite{Billault2021}.

The amplitude vectors $A_{{mn,d}}$ for each harmonic for each mode are sampled from uniform distributions, $A_{{mn,d}} \in \mathcal{R}(0.00,0.50)$.

\bibliography{article}

\end{document}